# Aharonov-Bohm effect in graphene

*Saverio Russo, Jeroen B. Oostinga, Dominique Wehenkel, Hubert B. Heersche, Samira Shams Sobhani, Lieven M.K. Vandersypen, and Alberto F. Morpurgo*[*]

Kavli Institute of Nanoscience, Delft University of Technology, PO Box 5046, 2600 GA, The Netherlands

[*]EMAIL: s.russo@tudelft.nl; a.morpurgo@tnw.tudelft.nl

ABSTRACT. We investigate experimentally transport through ring-shaped devices etched in graphene and observe clear Aharonov-Bohm conductance oscillations. The temperature dependence of the oscillation amplitude indicates that below 1 K the phase coherence length is comparable to or larger than the size of the ring. An increase in the amplitude is observed at high magnetic field, when the cyclotron diameter becomes comparable to the width of the arms of the ring. By measuring the dependence on gate voltage, we also observe an unexpected linear dependence of the oscillation amplitude on the ring conductance, which had not been reported earlier in rings made using conventional metals or semiconducting heterostructures.

The investigation of transport phenomena originating from quantum interference of electronic waves has proven to be a very effective probe of the electronic properties of conducting materials. Recent work has shown that this is also the case for graphene, a novel material consisting of an individual layer of carbon atoms, in which the electron dynamics is governed by the Dirac equation.[1] The anomalous behavior of the weak-localization correction to the conductivity that is observed in the experiments[2], for instance, is directly related to the presence of two independent valleys in the band structure of graphene[3,4], and can be used to extract the inter-valley scattering time.[5] Another example is provided by the observation of a Josephson supercurrent in graphene superconducting junctions, which permits to conclude that transport through graphene is phase coherent even when the material is biased at the charge neutrality point (i.e., where nominally no charge carriers are present).[6]

Possibly, the phenomenon that most directly illustrates electronic interference in solid-state devices is the occurrence of periodic oscillations in the conductance of ring-shaped devices, measured as a function of magnetic field.[7] This phenomenon, which is a direct consequence of the Aharonov-Bohm (AB) effect, has been investigated extensively in the past in rings made with metallic films or with semiconducting heterostructures, and its study has contributed significantly to our understanding of mesoscopic physics. For example, the analysis of h/e and h/2e AB conductance oscillations has clarified the difference between sample-specific and ensemble-averaged phenomena.[7] The investigation of the temperature and magnetic field dependence of the oscillation amplitudes has been used to investigate processes leading to decoherence of electron waves, such as electron-electron interaction[7,8], or the interaction with magnetic impurities.[9] In graphene, however, no experimental observation of AB conductance oscillations has been reported so far, although there is an emerging interest in the problem from the theoretical side.[10,11] In the course of recent experiments we have observed AB conductance oscillations experimentally in several rings fabricated on few-layer graphene. In this letter we report on systematic measurements that we have performed on a device made on single-layer graphene, as a function of temperature, density of charge carriers, and magnetic field.

The devices used in our experiments were fabricated following a by-now established procedure.[12] Thin graphite flakes were exfoliated from natural graphite using an adhesive tape, and transferred onto a highly-doped Si substrate (acting as a gate) covered by a 300 nm thick $SiO_2$ layer. The flakes were imaged under an optical microscope and their position was registered with respect to markers already present on the substrates. Single layer flakes were identified by looking at the shift of the light intensity in the green channel of the RGB scale relative to the adjacent substrate.[13] In subsequent nano-fabrication steps, the devices were patterned using electron-beam lithography to define pairs of metallic electrodes (Ti/Au), and to etch the rings in an Argon plasma. The rings have an inner and outer radius of 350 and 500 nm, respectively; the inset of Fig. 1a shows a scanning electron micrograph of a typical device.

Measurements were performed in a dilution refrigerator, in the temperature range between 150 and 800 mK. The conductance of the ring was measured using a lock-in amplifier, in a current-biased two-terminal configuration. For the different measurements the excitation current was varied to ensure that resulting voltage was smaller than the temperature, to prevent heating of the electrons and the occurrence of non-equilibrium effects.

Fig. 1a shows the resistance of the ring measured at T=150 mK, as a function of gate voltage. A clear peak, centered at approximately $V_G$= +4V, is observed as it is typical for graphene. The large peak resistance value may be expected since the rings are made of fairly narrow ribbons (~150 nm). It may originate either from the opening of a small gap due to lateral size quantization (as recently proposed[14]), or from the fact that, close to the charge neutrality point, disorder at the edges is not screened effectively, resulting in enhanced scattering with valley mixing and localization of electron states. To estimate the mobility μ of charge carriers we use the value of the conductance per square measured at high gate voltage. With the density of charge carriers being determined from the known capacitance to the gate (i.e., $G_{\square}=ne\mu=\varepsilon_0\varepsilon_r e\mu\ V_G/d$), we obtain μ=6000 $cm^2/Vs$, essentially independent of $V_G$ for $V_G>10V$ and $V_G<-10V$. The diffusion constant is estimated from the Einstein relation $\sigma=\nu e^2 D$, where σ is the measured conductivity and ν the density of states at the Fermi level, which for graphene is given by $\nu(\varepsilon_F)=g_v g_s 2\pi|\varepsilon_F|/(h^2 v_F^2)$.[15] Here $g_v$=2, $g_s$=2 account for the valley and spin degeneracy, $v_F=10^6$m/s is the Fermi velocity, and the value of $\varepsilon_F$ is determined by equating the expression for the charge density $n(\varepsilon_F)= g_v g_s \pi\varepsilon_F^2/(h^2 v_F^2)$ to the value determined by the gate voltage. We obtain D=0.06$m^2$/s, not far from the value of diffusion constant that is estimated assuming diffusive scattering at the ribbon edges (D=W $v_F$ = 0.15 $m^2$/s, with W width of the ribbon). With this value of the diffusion constant, we obtain a Thouless energy for the ring of $E_{Th}= \hbar D/L^2$ = 10 μeV (L is the ring circumference), which is slightly smaller than the lowest temperature (T=150 mK) at which the measurements have been performed.

The low-field magnetoresistance measured at $V_G$=30 V and T=150 mK is shown in Fig. 1b. The presence of periodic oscillations is clearly visible. The period in field is approximately ΔB= 7 mT and the corresponding Fourier spectrum is shown in Fig. 2b. In the spectrum a peak is present, whose position and width correspond well to what is expected for h/e oscillations given the values of the inner and outer radius in our device. Fig. 2a shows the evolution of the AB conductance oscillations (with the background removed by subtracting the magnetoresistance averaged over one period of the oscillations) measured at different temperatures. It is apparent that their amplitude decreases with increasing temperature. This decrease is quantified in the inset of Fig. 2b, where we plot the root mean squared value of the amplitude as a function of T, in a double logarithmic scale. We find that the oscillation amplitude is proportional to $T^{-1/2}$. This dependence, which is commonly observed in metal rings, is due to thermal averaging of the h/e oscillations ($\delta G_{AB} \propto (E_{TH}/k_B T)^{1/2} \exp(-\pi r/L_\phi(T))$, where r is the radius of the ring)[7]. It is expected for temperature values larger than the Thouless energy (which is the case here), if the phase coherence $L_\varphi = (D\tau_\varphi)^{1/2}$ length is longer than the arms of the ring. Indeed, with the value of diffusion constant given above and taking for the phase coherence time values estimated in the literature (away from the charge neutrality region $\tau_\phi$ ~ 0.1 ns at T=1K, increasing roughly linearly with decreasing T)[16], we find that in our ring also this condition is satisfied. A similar $T^{-1/2}$ dependence was observed for all gate voltages at which the AB oscillation amplitude was sufficiently large to be accurately measured.

The detailed dependence of the amplitude of the AB conductance oscillations on gate voltage is particularly interesting. It is apparent from Fig 3a and b that at the charge neutrality point essentially no AB oscillations are observed and that the amplitude of the oscillations becomes larger as the charge density in the sample is increased. This finding is summarized in Fig. 3c, in which the dependence of rms amplitude on $V_G$ is shown to anti-correlate with the gate-voltage dependence of the total sample resistance. Indeed, if we plot the amplitude of the AB oscillations as a function of the device conductance (Fig 3d), we observe that a linear relation is surprisingly well obeyed. Such a relation had not been observed previously in metallic rings, nor in rings formed in semiconducting heterostructures, and it is only predicted theoretically for rings containing tunnel barriers.[17]

All measurements discussed so far have been performed at magnetic fields smaller than 0.5 T, for which the maximum observed amplitude of the AB conductance oscillations is only 0.02 $e^2/h$. In Fig. 4c we now plot the AB conductance oscillations measured in a larger magnetic field range (up to 9 T): the data clearly show that the rms amplitude of the oscillations is significantly larger at higher fields than around B=0T (see also Fig. 4a and Fig. 4b). We have performed a quantitative analysis of the evolution of the h/e oscillations by determining their rms amplitude in a 350 mT interval (approximately 50 periods) as a function of magnetic field. Figure 5a shows that, irrespective of the temperature at which the measurement is performed, the oscillation amplitude increases and saturates starting from approximately 3T. The relative increase is also comparable in magnitude (roughly 4 times) at the two different temperatures. Owing to the large amplitude, in the high field regime the second harmonic in the Fourier spectrum of the AB oscillations becomes visible (see Fig. 4d, obtained from the measurement of magnetoresistance between 4 and 5 Tesla). The temperature dependence of the amplitude of both the h/e and h/2e component measured at high field still scales linearly with $T^{-1/2}$, similarly to what is seen in the low field regime.

An increase in the amplitude of the AB conductance oscillations with increasing magnetic field is unusual. In metallic rings, a known mechanism causing such an increase is scattering off magnetic impurities[7,9]. At low field the spin of these impurities can flip without any energy cost, causing dephasing of the electron waves. At higher field, the finite Zeeman energy prevents the spins to flip, leading to a decrease in dephasing and a corresponding increase in AB oscillation amplitude. Despite the fact that magnetic impurities have been predicted to form at defects in graphene or could be present at the edges[18,19], in our device the presence of magnetic impurities cannot account for the experimental observations. In fact, if the effect observed was due to spin, one should observe that the magnetic field required for the enhancement of the oscillation amplitude increases with temperature (since the Zeeman energy has to be larger than $k_BT$), which is not what we see. An indication as to the origin of the anomalous increase observed in our graphene ring comes from the estimate of the diameter of the cyclotron orbit. At $V_G$=30V and 3 Tesla –the field at which the amplitude starts saturating- the cyclotron diameter is approximately 140 nm, comparable to the width of the ribbons forming the AB ring. This suggests that the effect of the field is of orbital nature. Indeed, Fig. 5b shows the amplitude of the AB conductance oscillations as a function of field measured at $V_G$=20V, where the increase occurs for a slightly smaller magnetic field, as it is expected since the cyclotron diameter ($hk_F/\pi eB$) is smaller at lower carrier density. The precise nature of the orbital mechanism leading to larger oscillation amplitude at higher magnetic field remains to be determined, but we suspect that the phenomenon originates from an asymmetry present in the arms of the ring caused by defects or inhomogeneity in graphene[20].

Interestingly, the linear relation between the amplitude of AB oscillations and the device conductance persists also at high field. Although less data are available -we have only performed high field measurements at VG = +4, +20 , and +30 V-, when plotted as a function of conductance the points still fall on one single line (see Fig. 5d). As mentioned above, such a linear relation is only predicted theoretically for rings containing tunnel barriers in their arms[17]. This may indicate that, consistent with the scenario that we propose to explain the observed magnetic field dependence[20], the graphene ring is rather strongly inhomogeneous and contains small, highly resistive regions acting as weak links or tunnel barriers. Whereas the presence of such highly resistive regions would not be surprising in the low density regime, it is less obvious that it should be expected when the density of charge carriers is of the order of 2-5 $10^{12}$ $cm^{-2}$ (corresponding to the high $V_G$ values in our experiments). A theoretical analysis of the relation between AB oscillation amplitude and ring conductance in graphene devices is called for, and it is likely that it will prove insightful to understand the nature of transport through narrow graphene ribbons.

In conclusion, we have reported the first observation and systematic study of Aharonov-Bohm conductance oscillations through a graphene ring. We find that in rings with a diameter of approximately 1µm, the phase coherence length of electrons is comparable to or longer than the device size for temperatures below 1 K. As a result, the oscillation amplitude increases as $T^{-1/2}$ with decreasing temperature, owing to thermal averaging on an energy scale larger than the Thouless energy. We also observe an anomalous magnetic field dependence of the AB oscillation amplitude, originating from an orbital effect of the magnetic field. Surprisingly, measurements as a function of gate voltage show that the amplitude of the conductance oscillations scales linearly with the total conductance of the device, a phenomenon that had not been previously observed in AB rings made using metal films or semiconducting heterostructures.

**Acknowledgment.** We thank I. Adagideli, Y. Blanter, and Yuli V. Nazarov for useful discussions, M.F. Craciun, X. Liu, T. Meunier, K.C. Nowack, and I.T. Vink for help in the experiments, and T.M. Klapwijk and L.P. Kouwenhoven for allowing us to use equipment in their laboratory. Financial support is obtained from the Dutch Organization for Fundamental Research on Matter (FOM), the 'Netherlands Organization for Scientific Research' (NWO-VICI program), and NanoNed.

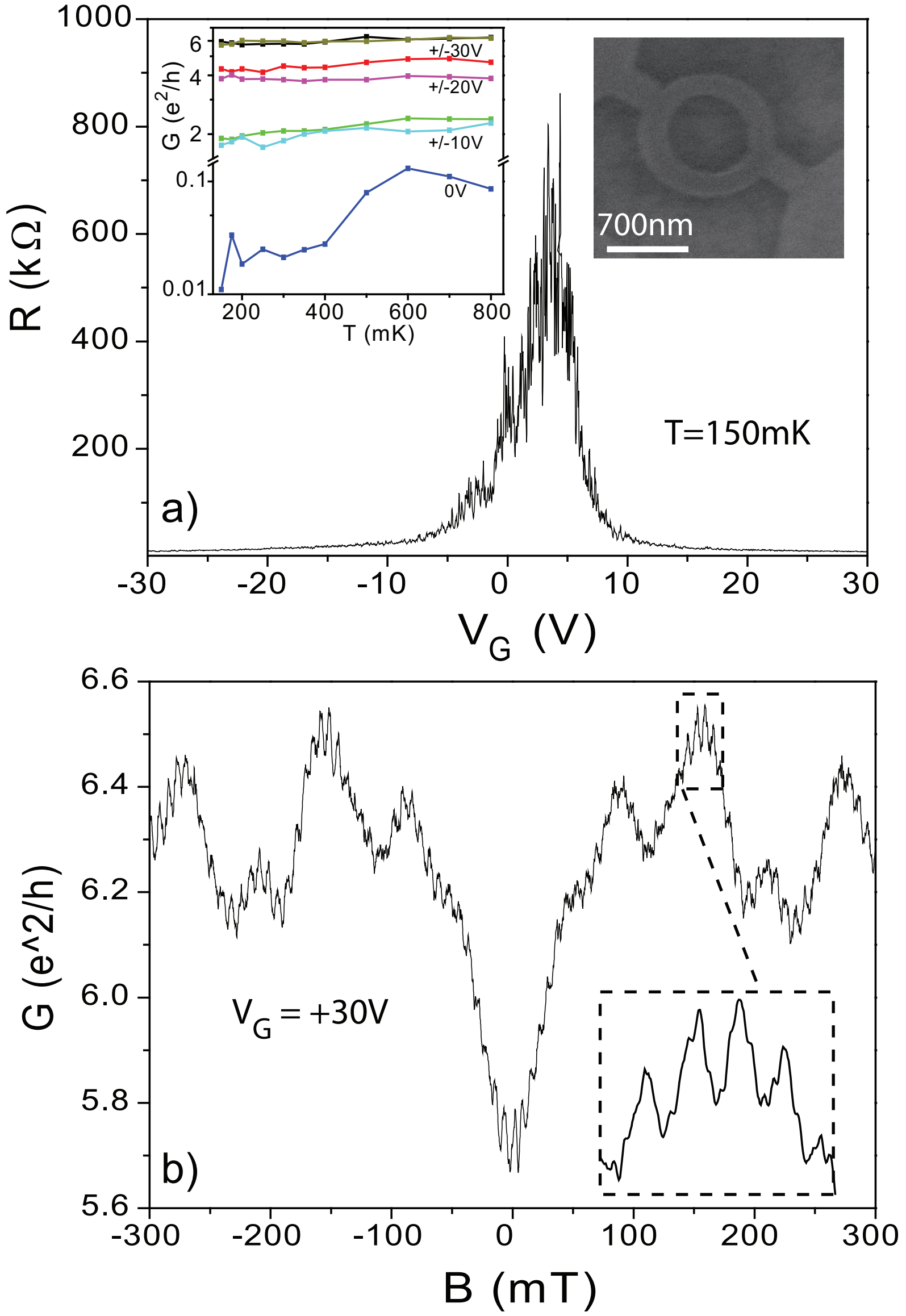

**Figure 1.** (a) The main panel shows the two-probe measurement of the ring resistance versus back gate voltage at T=150mK. The charge neutrality point is at +4V. Left inset: temperature dependence of the conductance measured for different values of gate voltage. Right inset: SEM image of a ring-shaped device etched in graphene similar to the one used in our measurements. (b) Magnetoconductance of the graphene ring measured at T=150mK and $V_G$=+30V. On top of the aperiodic conductance fluctuations, periodic oscillations are clearly visible as also highlighted in the inset.

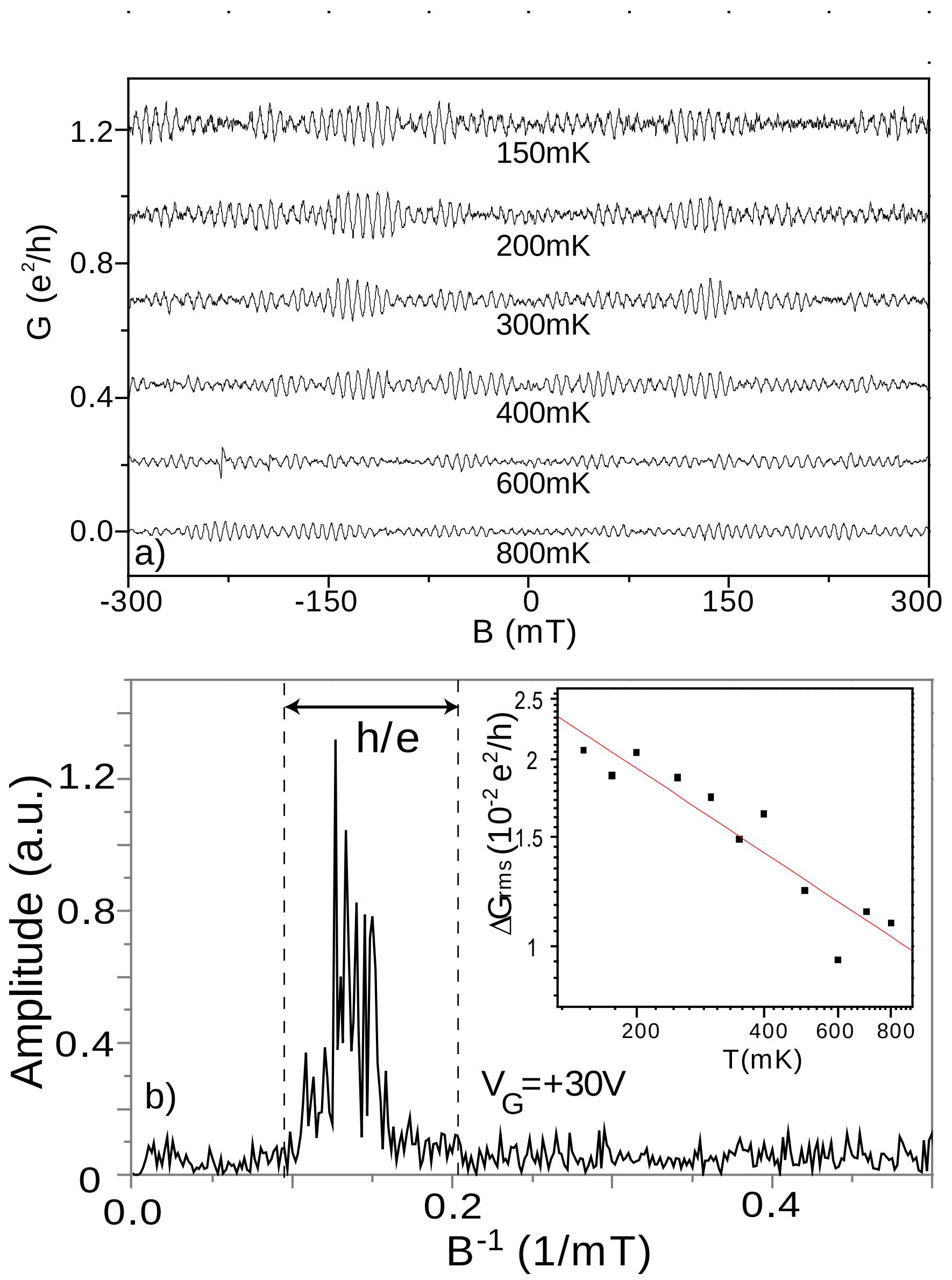

**Figure 2. (**a) Aharonov-Bohm conductance oscillations measured at $V_G$=+30V for different temperatures (curves are offset for clarity). (b) Fourier spectrum of the oscillations measured at T=150 mK shown in panel (a). The vertical dashed lines indicate the expected position of the h/e peak, as determined from the inner and outer radius of the ring. In the inset: temperature dependence of the root mean squared amplitude of the AB conductance oscillations, obtained from the measurements shown in panel (a).

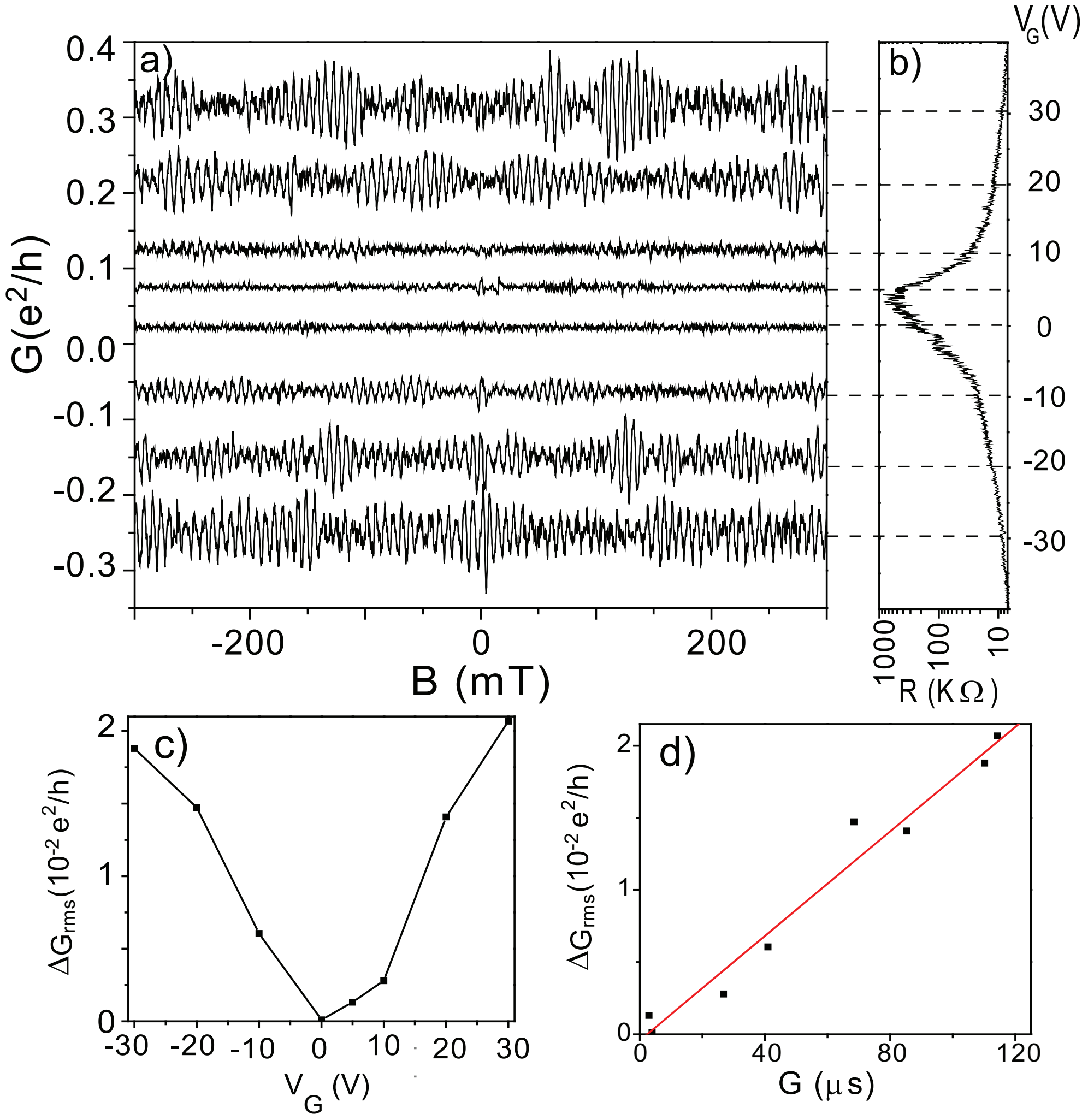

**Figure 3.** (a) AB conductance oscillations measured at T=150 mK, for different values of the back gate voltage, as indicated by the dashed lines on panel (b). Panel (b) shows the dependence of the ring resistance (in logarithmic scale) on gate voltage, measured at T=150 mK. (c) rms amplitude of the AB conductance oscillations as a function of gate voltage. In panel (d) the same data is plotted as a function of the ring conductance. The red line is a guide to the eye.

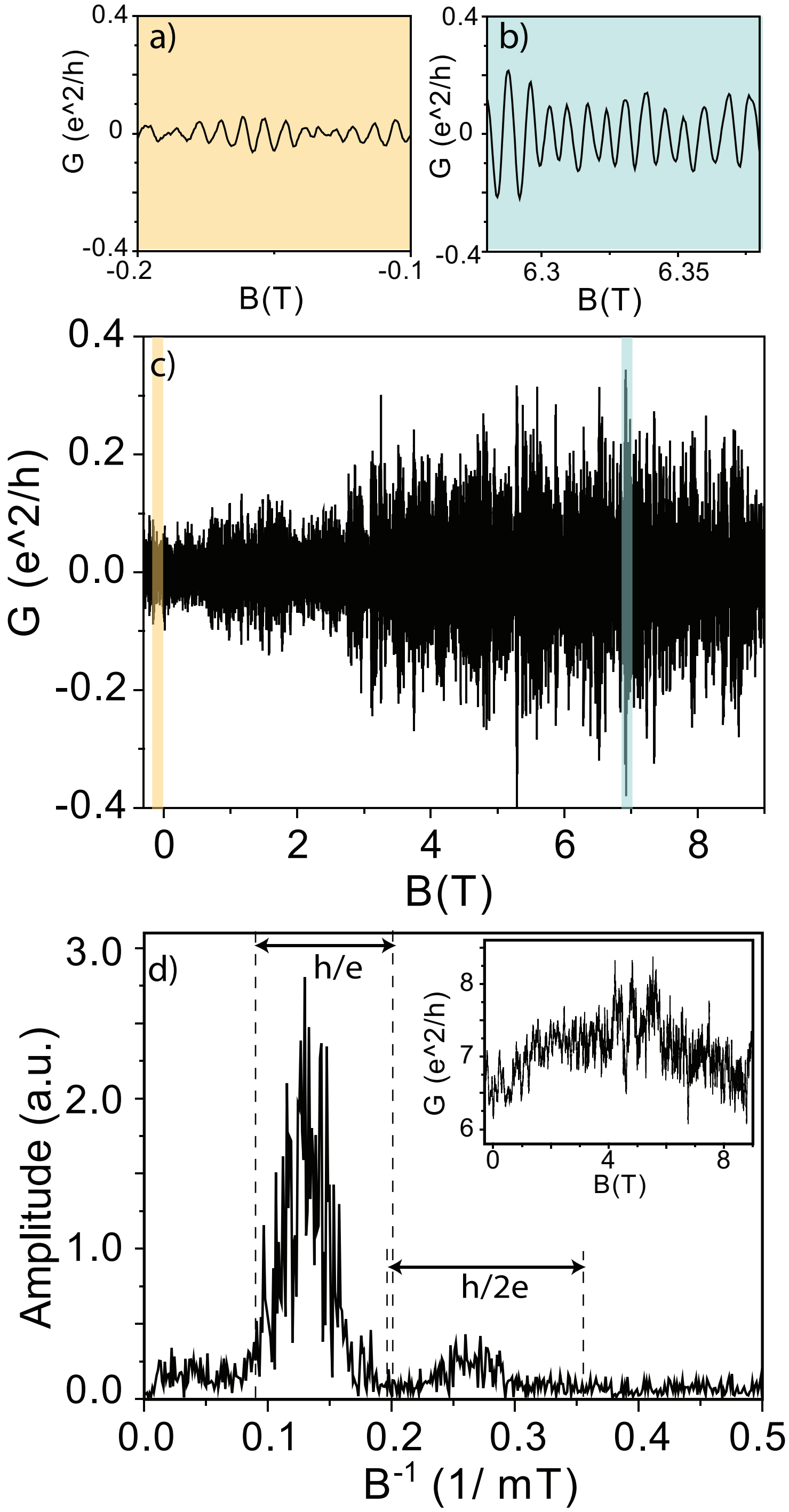

**Figure 4.** Panels (a)-(c) show the AB conductance oscillations measured at $V_G$=+30V in different magnetic field ranges. For B~3T a clear increase of AB amplitude is observed. (d) Fourier spectrum of the AB oscillations measured between 4T and 5T. The dashed lines indicate the position of the h/e and h/2e peaks expected from the device geometry. The inset shows the magnetoconductance of the ring. All measurements were taken at 150 mK.

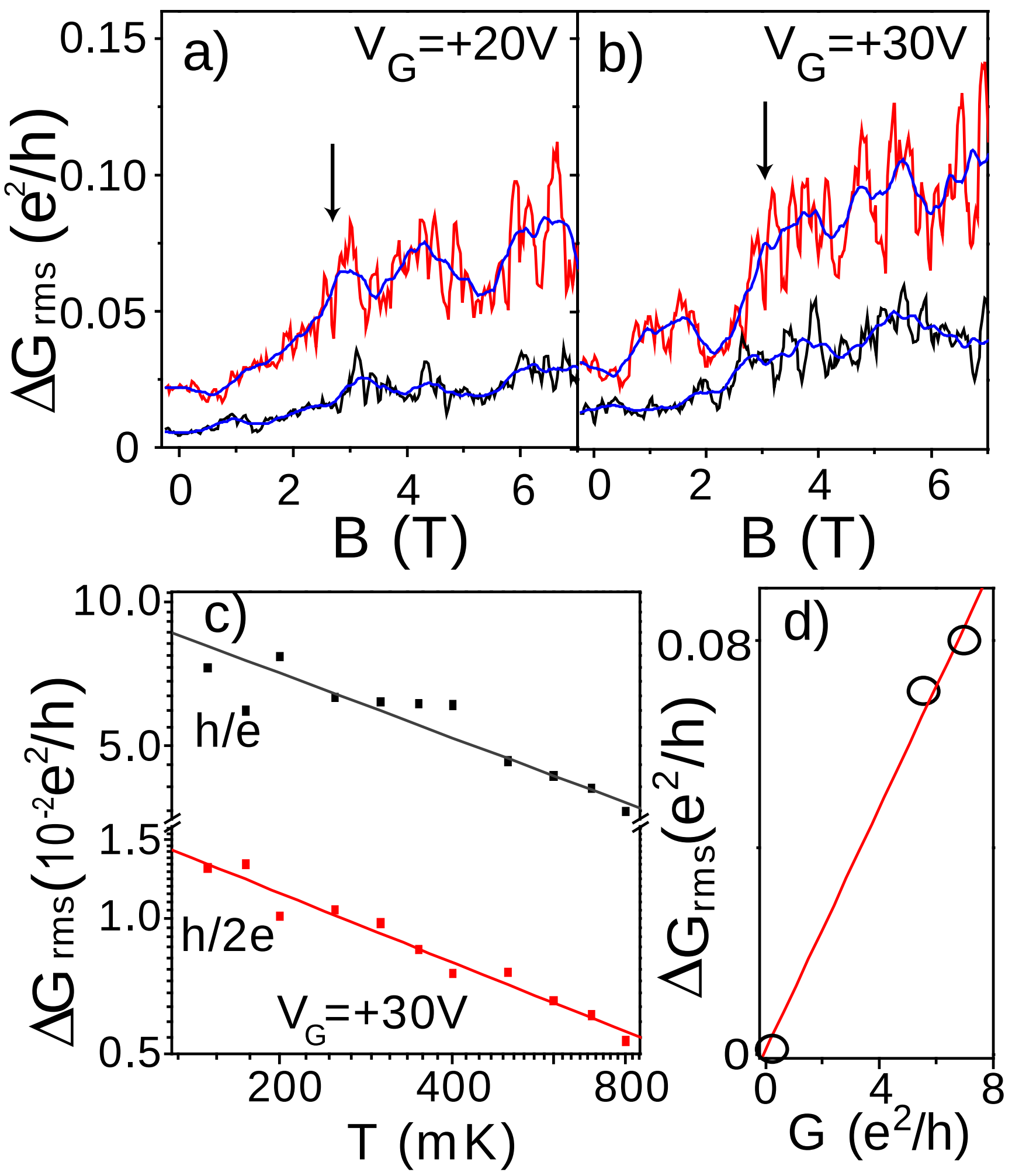

**Figure 5.** Panels (a) and (b) show the rms amplitude of the AB conductance oscillations determined on an interval of 50 periods, as a function of magnetic field and for two different temperatures (T=150mK, red line; and 800mK, black line), measured at $V_G$=+20 and +30 V. The blue curves are obtained by smoothing the corresponding data over 20 points. The arrows indicate the position in field where the oscillation amplitude starts to saturate, slightly higher at $V_G$=+30. Panel (c) shows the temperature dependence of the amplitude of the h/e and h/2e components of the AB conductance oscillations measured between 4 T and 5T. The continuous lines are linear fits with slopes -0.5±0.07. Panel (d) shows that the rms values of the AB oscillations measured between 4 and 5 T (at $V_G$ = +4, +20, +30V) scales linearly with the conductance of the device.